\documentclass[preprint,12pt]{elsarticle}

\usepackage{amssymb}
\usepackage{amsmath}

\journal{Nuclear Physics A}

\begin{document}

\begin{frontmatter}



\title{Isospin Dynamics and Dipolar degrees of freedom in Heavy Ion collisions - Study of the system $^{48}$Ca+$^{27}$Al at 40 MeV/nucleon} 

\author[1]{M. Papa \corref{cor1}}
\author[1]{G. Cardella}
\author[2]{L. Acosta}
\author[1]{E. De Filippo}
\author[3]{F. Favela}
\author[1,4]{E. Geraci}
\author[1,4]{B. Gnoffo}
\author[5]{C. Maiolino}
\author[1]{N.S. Martorana}
\author[1]{A. Pagano}
\author[5]{E. V. Pagano}
\author[1]{S. Pirrone}
\author[1,4]{G. Politi}
\author[1,6]{L. Quattrocchi}
\author[1,6]{F. Risitano}
\author[4,5]{F. Rizzo}
\author[5]{P. Russotto}
\author[1,6]{A. Trifir\'{o}}
\author[1,6]{M. Trimarchi}
\author[5,7]{C. Zagami}
\cortext[cor1]{Correspondng author}

\affiliation[1]{organization={Istituto Nazionale di Fisica Nucleare -Sezione di Catania},
		  addressline={Via S.Sofia 64}, 
             city={Catania},
            postcode={95123}, 
            state={Italy}}
\affiliation[2]{organization={Instituto de Estructura de la Materia, CSIC}, city={Madrid},state={Spain}}
\affiliation[3]{organization={Instituto de Fi­sica. Universidad Nacional Autonoma de Mexico}, city={Ciudad de Mexico},state={Mexico}}
\affiliation[4]{organization={Dipartimento di Fisica e Astronomia ”Ettore Majorana”}, city={Catania},state={Italy}}.

\affiliation[5]{organization={Istituto Nazionale di Fisica Nucleare, Laboratori Nazionali del Sud}, city={Catania},state={Italy}}.

\affiliation[6]{organization={Dipartimento di Scienze MIFT, Univ. di Messina}, city={Messina},state={Italy}}.

\affiliation[7]{organization={CSFNSM - Centro Siciliano di Fisica Nucleare e Struttura della Materia}, city={Catania},state={Italy}}.

\begin{abstract}
The isospin-equilibration process is investigated in the reaction 
$^{48}$Ca+$^{27}$Al at 40 MeV/nucleon by means of the $4\pi$ multi-detector CHIMERA through the dipolar observable 
$\langle V_c \rangle$, constructed from the momenta and the charge/mass 
asymmetry of the produced fragments and particles. The experimental 
results show a reduction of the dipolar signal with increasing 
dissipation, indicating a progressive approach toward the 
equilibration.
\noindent
For the selected processes, the comparison between experimental 
data and CoMD-II calculations indicates that this observable is 
sensitive to the symmetry energy over a range of average densities 
from about the saturation value $\rho_{0}$ up to about $1.6\rho_{0}$. 
A likelihood analysis constrains the symmetry energy at saturation 
to $E_{\text{sym}}(\rho_{0}) = 32. \pm 2$ MeV and yields a slope parameter 
$L = 68 \pm 10$ MeV and $L = 73 \pm 10$ MeV for incompressibility module equal to 220 and 260 respectively. 
From the same analysis we can also reliably estimate
$E_{\text{sym}}(1.2\rho_{0})=37\pm 4$ MeV.
\end{abstract}




\end{frontmatter}



\section{Introduction}
One of the primary goals of heavy-ion collision (HIC) studies is to constrain the nuclear matter equation of state (EoS) over a range of densities different from the saturation. This information is of particular relevance because the symmetry energy plays a key role in determining the properties of neutron-star matter, including neutron-star mergers and core-collapse supernovae \cite{lat1,lat2,stein,rob,abb}.

\noindent
Isospin equilibration in HIC involving nuclei with different charge-to-mass asymmetries provides direct access to the symmetry energy. In the hydrodynamical limit \cite{bal,bar}, the neutron-to-proton flux can be decomposed into two contributions driven by density and concentration gradients, corresponding to isospin drift and diffusion, respectively. The associated transport coefficients depend explicitly on the symmetry energy and on its density derivative \cite{bal}.
\noindent
Most experimental investigations of isospin equilibration have focused on diffusion effects and migration of the neutron excess through the neck region. Systems with different projectiles (P) and targets (T) asymmetries have been studied around the Fermi energy, and observables related to isotopic yields of produced clusters have been analyzed through the Isospin Transport Ratio \cite{rami,tsang,lynch,fable,piant}.
As discussed in Section 2, it should be noted, however, that the relative collective motion of light particles,in particular, emitted neutrons and protons, also contributes to the equilibration process trough the related fluxes\cite{Bao,russo1}.
\noindent
A different and comprehensive approach, is based on the measurement of the so-called dipolar signal obtained from the momenta and the charge/mass asymmetry of the produced fragments and particles. In particular, this observable \cite{JPC,diprap} is directly connected to the early dynamical stage of the reaction and to the collective isovector dynamics preceding the statistical decay phase.

\noindent
In this work, we investigate the system $^{48}$Ca+$^{27}$Al at 40 MeV/nucleon by means of calculations performed with the Constrained Molecular Dynamics model CoMD-II \cite{comd2,comd3}. We analyze the sensitivity of the dipolar observable to the degree of dissipation collision, to the symmetry energy values and its stiffness parameter. Particular attention is devoted to the connection between the time evolution of the dipolar signal and the density evolution during the collision.

\noindent
Section 2 introduces the dipolar observable and discusses its relation to pre-equilibrium $\gamma$-ray emission and to the isospin equilibration dynamics. In Section 3 we present experimental results obtained with the CHIMERA multi-detector \cite{pag,card} at the Laboratori Nazionali del Sud (Catania, Italy) and the main criteria of data analysis. Section 4 contains the comparison between data and model calculations. Section 5 summarizes the conclusions.

\section{The dynamical Isospin equilibration}
\subsection{Dipolar degrees of freedom and the Isospin equilibration processes}
In this section, we establish a simple connection between dipolar degrees of freedom in HIC and the isospin equilibration process.

\noindent
It is well known that the excitation of dipolar degrees of freedom in nuclei is associated with one of the most widely studied collective motions: the Isovectorial Giant Dipole Resonance (GDR) \cite{snover,gaard}. As with other electromagnetic excitations, the GDR can be built upon any excited state. Its possible persistence (i.e., finite damping) even at high temperatures has been extensively investigated \cite{sant}.

\noindent
According to the droplet model, this mode can be described macroscopically as harmonic oscillations of the neutron sphere with respect to the proton sphere (GT mode), and/or as out-of-phase density oscillations of neutrons and protons inside a fixed spherical volume (JS mode) \cite{myers}. For a nucleus of mass $A$, radius $R$, with $N$ neutrons and $Z$ protons, the excitation is associated with a dipole moment (expressed in units of the elementary charge):

\begin{equation}
\textbf{D}_{j} = \alpha_{j} \dfrac{NZ}{A} \textbf{R}
\label{eq6}
\end{equation}

The coefficients $\alpha_{j}$ determine the strength of the two possible modes.
\noindent
The possibility of exciting exotic GDR modes
during the formation of intermediate systems 
produced by the collision of nuclei with different
\noindent 
asymmetries $\beta = \dfrac{N-Z}{A}$ was investigated starting from the early 90s. 
\noindent
An excess of gamma-ray yield around 9 MeV, beyond what is predicted by statistical models \cite{brink}  was observed in several systems at incident energies between 9 and 25 MeV/nucleon \cite{amo,papa1,pierrou,paras0,paras1}. These observations have been linked to pre-equilibrium dynamical dipole emissions.

\noindent
For systems composed of partners with different $\beta$ values, and at the touching configuration, the dipolar degrees of freedom can be described in the center-of-mass system (CM) as:

\begin{equation}
 \textbf{D}  =  \dfrac{1}{2} \mu (\beta_T - \beta_P) \textbf{R} , \quad
 \textbf{V}  =  \dfrac{1}{2} \mu (\beta_T - \beta_P)(\textbf{V}_P - \textbf{V}_T) 
\label{eq2}
\end{equation}

Here, $\mu$ is the reduced mass number of the binary system formed by the projectile-like and target-like fragments, $\textbf{R}$ is their distance, and $\textbf{V}_P$, $\textbf{V}_T$ are their velocities. $\textbf{D}$ represents the dipole moment due to the neutron-proton center-of-mass separation.
Assuming that $\mu$ and asymmetries $\beta$ remain nearly constant at the onset of the interaction, the time derivative of $\textbf{D}$ is given by $\textbf{V}$. Its time variation leads to pre-equilibrium $\gamma$-ray emission. In a simple binary picture, the expression for  $\textbf{V} $ remains valid using final state values of the projectile and target like fragments. If charge and mass transfer reach equilibrium, then $ \beta_P^f \simeq  \beta_T^f $, and thus $ \textbf{V}$ becomes negligible, this in our description defines the dynamical isospin equilibration process. In the following 
$\textbf{V}$ will be called dipolar signal.
\noindent
Due to the linear dependence from the fragment velocities, in more complex scenarios $\langle \textbf{V} \rangle$ can be decomposed as:
\[
\langle \textbf{V} \rangle = \langle \textbf{V}_G \rangle + \langle \textbf{V}_{LG} \rangle + \langle \textbf{V}_L \rangle
\]
where $\textbf{V}_G$ and $\textbf{V}_L$ are the average dipolar signals from the "gas" (light charged particles) and "liquid" (main fragments), respectively. $\textbf{V}_{LG}$ refers to the contribution associated to the relative motion between these two phases.
Hereafter, the curly brackets generally denote averages over event ensembles selected according to other quantities or observables characterizing the system under study.

\noindent
Assuming the gas phase consists of neutrons and protons, we further express:

\begin{equation}
\langle \textbf{V} \rangle =
\left\langle \dfrac{A_g(1 - \beta_g^2)}{4} \textbf{V}_{pn} \right\rangle +
\left\langle \mu_{lg} \dfrac{\beta_l - \beta_g}{2} \textbf{V}_{lg} \right\rangle +
\langle \textbf{V}_L \rangle
\label{eq3}
\end{equation}

\noindent
In the above expression the first term represents the contribution related to the proton-neutron relative motion of the gas ”phase” composed of $A_{g}$ nucleons, expressed through the relative velocity $\textbf{V}_{pn}$ i.e. it represents the associate average differential flux; the second term concerns the relative velocity $\textbf{V}_{lg}$ between the centers of mass of the ”liquid” fragments and the ”gas”, $\mu_{lg}$  is the reduced mass associated to the two sub-systems and the last term  represents the contribution produced by the relative motion of the main fragments. A similar expression can be obtained including others light particles in the gas ”phase”.
From this decomposition we observe that  the dynamical equilibration condition $\langle\textbf{V}\rangle\simeq 0$, for the total system, requires a delicate balance which depends on the average neutron excess of the produced ”liquid drops”, the one associated to the ”gas” pre-equilibrium emission($\beta_{g}$), and on the relative velocities between the different parts. To enlighten the role played by some of the terms reported in Eq.(3), we can consider the idealized decay of one  source through neutrons and protons emission. This example can also schematically describe the case of the complete stopping in heavy ion collisions. Moreover, for simplicity, we can consider uncorrelated fluctuations between the velocities, masses and neutron excesses. For non-identical colliding nuclei, if pre-equilibrium emission exists, then $\langle\textbf{V}_{lg}\rangle\neq 0$. Because we are supposing the formation of one main source $<\textbf{V}_{L}>$=0. In this case, if $\langle \beta_{g}\rangle\neq \langle \beta_{l}\rangle$ and $\langle \beta_{g} \rangle \ne 1$, due, for example, to the isospin  ”distillation” phenomenon, to obtain
the dynamical equilibrium condition $\langle\textbf{V}\rangle=0$
the first term of Eq. 3 has to be necessarily different from zero and it will contribute to the neutron-proton differential flow.
This indicates that understanding the global isospin equilibration dynamics requires accounting for the pre-equilibrium "gas" contribution.

\noindent
A more realistic treatment of course requires a full many-body approach.

\noindent
In the following  we can also estimate the energy $\Delta E$ associated with these degrees of freedom as:

\begin{equation}
\Delta E = A_P E_\text{{sym}} \beta_P^2(t) + A_T E_\text{{sym}} \beta_T^2(t) - A E_\text{{sym}} \beta^2(t) + \dfrac{1}{2} \mu_{NZ} v_{\text{rel}}^2
\label{eq4}
\end{equation}

Here:

- $\beta_P$, $\beta_T$, and $\beta$ are the charge/mass asymmetries of P, T, and of the total system.

- $E_{\text{sym}}(\rho_{0})$ is the symmetry energy per nucleon at the saturation density.

- $\mu_{NZ}$ is the reduced mass corresponding to all the neutrons and protons.

- $v_{\text{rel}}$ is the relative velocity between neutron and proton centers of mass, approximated by $\textbf{v}_{\text{rel}} = \left( \dfrac{N_P}{N} - \dfrac{Z_P}{Z} \right) \textbf{V}_b$.

\noindent
$\textbf{V}_b$ is the beam velocity through
which for a given system  the excitation of the pre-equilibrium mode can be enhanced.
Apart from small differences due to the Coulomb and surface contributions, $\Delta E$ represents the total energy due to dipolar mode, compared to a reference binary system having the same total mass $A$ and charge $Z$, and equal charge/mass ratios.

\noindent
For instance, in the case of $^{48}$Ca + $^{27}$Al at 40 MeV/nucleon, the symmetry energy contribution is about 8.7 MeV (assuming $E_{\text{sym}}(\rho_{0}) = 30$ MeV), and the kinetic part from the beam energy is about 2.6 MeV.

\noindent
Despite the approximations in Eq.~(\ref{eq4}), its structure highlights key factors influencing the dipolar time evolution: drift effects (from symmetry energy density gradients) and diffusion effects between the partners (from spontaneous and temperature-dependent $\beta$ changes). But dealing with  open systems there will be also significant outgoing contributions.

\noindent
For a given colliding system and a dynamical microscopic model
the time evolution of ensemble-averaged dipolar signal $\langle \textbf{V} \rangle$ can be evaluated  from
the microscopic configurations. That is:
 $\langle \textbf{V} \rangle=\langle\sum_{i=1}^{Z} \textbf{v}_{i}(t)\rangle$  
with $\textbf{v}_i$ being the velocities of protons in the CM.
 The average pre-equilibrium  component along the impact parameter is weak, while the one normal to the reaction plane is not excited due to symmetry. Large effect can be obtained for the component along the beam direction $\langle V \rangle$.
More generally, as shown in Appendix B of Ref. \cite{papa1}, the late statistical decay of hot fragments does not affect the average of total dipolar signal due to conservation laws and symmetry of this kind of decay processes. Therefore, asymptotically in absence of non-statistical dipolar excitation of the fragments, $\langle \textbf{V} \rangle$ can be related to the final fragmentation as:
\begin{equation}
\langle \textbf{V} \rangle = \dfrac{1}{m_0} \sum_{Z,A,k=1}^{n} \dfrac{Z}{A} \dfrac{m_{Z,A}^k \overline{\textbf{P}}_{Z,A}^k}{n}
= \dfrac{1}{m_0} \sum_{Z,A} \dfrac{Z}{A} \langle m_{Z,A} \rangle \langle \overline{\textbf{P}}_{Z,A} \rangle C_{Z,A}
\end{equation}
Here, $n$ is the number of events, $\overline{\textbf{P}}_{Z,A}^k$ is the average momentum of fragments with charge $Z$ and mass $A$ in event $k$, $m_{Z,A}^k$ is the multiplicity, and $C_{Z,A}$ is the correlation function:
\begin{equation}
C_{Z,A} = \dfrac{\langle m_{Z,A} \, \overline{\textbf{P}}_{Z,A} \rangle}{\langle m_{Z,A} \rangle \langle \overline{\textbf{P}}_{Z,A} \rangle}
\end{equation}
This shows that $\langle \textbf{V} \rangle$  can be related to the final average charge/mass distribution weighted in the momentum space of the fragments via correlations between multiplicities and momenta.

\noindent
For the investigated system $^{48}$Ca + $^{27}$Al at 40 MeV/nucleon
we performed CoMD-II calculations with an effective Hamiltonian  derived from a local Skyrme interaction with a symmetric nuclear matter incompressibility equal to $K_{0}$=220 and 260 MeV.
Fig.~1 shows the asymptotic value of the beam axis component $\langle V \rangle$ of the dipolar signal as a function of the impact parameter for two extreme values of the stiffness parameter $\gamma$ entering in the symmetry potential per nucleon. The symmetry energy at density $\rho$ is written within the mean-field approximation, as\cite{comd3} :
\begin{equation}
E_{\text{sym}}(\rho)=V_{\text{sym}}(\rho)+\dfrac{1}{3}E_{F}(\rho_{0})(\dfrac{\rho}{\rho_{0}})^{2/3} \qquad V_{\text{sym}}(\rho) = V_0 \left( \dfrac{\rho}{\rho_0} \right)^\gamma
\end{equation}
where $E_{F}(\rho)$ is the Fermi energy a the saturation density.
The figure reveals the sensitivity of the dipolar observable to the $\gamma$ parameter. We also observe that the degree of dynamical isospin equilibration $R_{eq}$ reach the larger values for intermediate parameters, with the super-soft option ($\gamma = 0.5$) being the most equilibrated. For $b=4 fm$ by considering an initial value of about -9.8 cm/nsec we obtain $R_{eq}\simeq 72$\%.
\begin{figure}[h]
\centering
\includegraphics[width=10cm,clip]{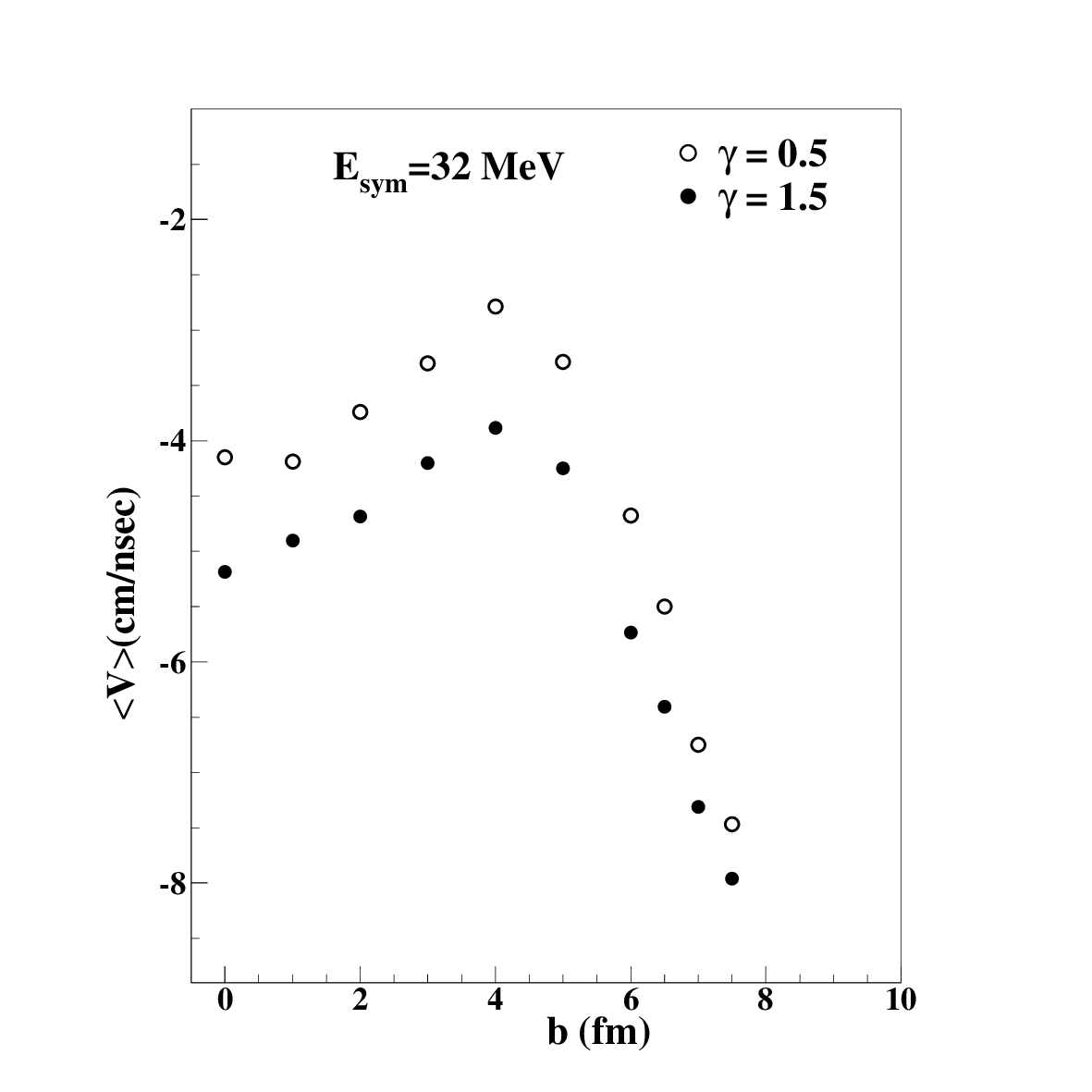}
\caption{For $K_{0}$=220 MeV, $\langle V \rangle$ as a function of the impact parameter $b$, obtained from CoMD-II calculations. The values of the symmetry energy and of the stiffness parameters $\gamma$ are indicated in the figure.}
\end{figure}

\subsection{Dipolar signal and the sensitivity to the change of the density}
A central question concerns the identification of the density region effectively probed by a given isospin-sensitive observable. In heavy-ion collisions, the density evolves rapidly and spans both sub-saturation and supra-saturation values within a short time interval. Therefore, the sensitivity of an observable to the symmetry energy can not be associated to a well-defined density.

This ambiguity is not limited to the dipolar signal but also applies to other observables. A quantitative analysis of the correlation between the time evolution of the observable and the corresponding density evolution is thus required in order to clarify which density regions predominantly determine the asymptotic signals.

\begin{figure}[h]
\centering
\includegraphics[width=10cm,clip]{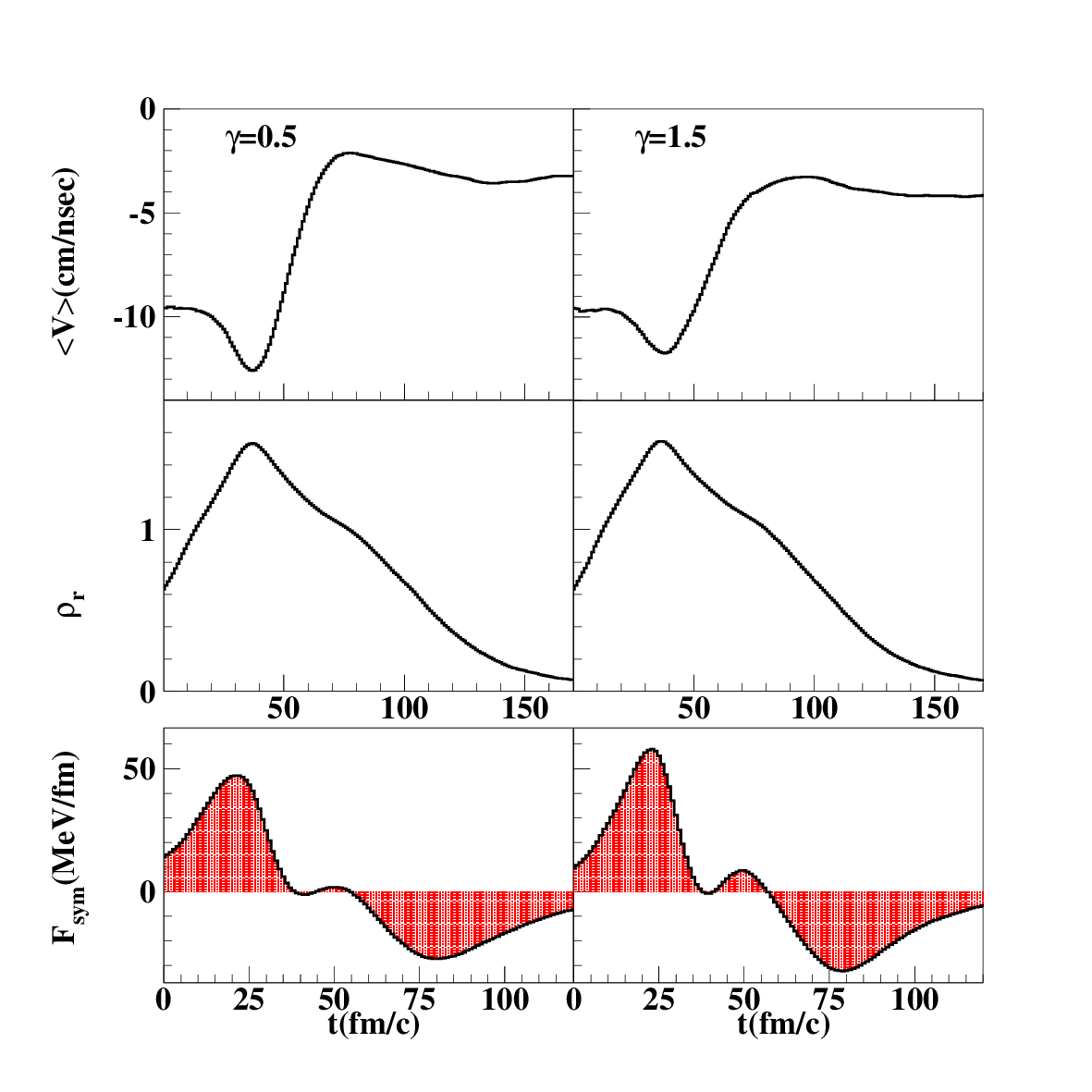}
\caption{(Colour on-line)
Results of CoMD-II calculations for $E_{\text{sym}}(\rho_{0})=32$ MeV, impact parameter $b=3$ fm, and different values of the stiffness parameter $\gamma$, shown as a function of time. Top panels: $\langle V \rangle$ (see also Fig. 1 for the asymptotic value). Middle panels: local relative density around the center of mass. Bottom panels: average local force between protons and neutrons. The incompressibility $K_{0}$  is equal to 220 MeV}
\end{figure}
For this purpose, as an example, Fig.~2 (upper panels) shows the time evolution of the average dipolar signal obtained from CoMD-II calculations
for an impact parameter $b =3$ fm, symmetry energy $E_{\text{sym}}(\rho_{0}) = 32$ MeV, and the two extreme values of the stiffness parameter $\gamma$
\footnote{The present study, according to the prescription given in Ref\cite{comd2}, includes accurate preliminary calculations to select the initial configurations with regulation of the surface term. Such a procedure is required in the constrained molecular dynamics to simulate stable nuclei over a time interval larger than 100 fm/c while preserving binding energies and root-mean-square radii within 10–15\% of the experimental values.} 
The value of the selected impact parameter corresponds approximatively to the average impact parameter evaluated for the most dissipative events analyzed in this work (see next sections). The study in time of $\langle \textbf{V}\rangle$ allow to defines the pre-equilibrium time interval over which the dipolar signal approaches its asymptotic value which is accessible experimentally through fragment detection. The difference in the asymptotic value are better evidenced by looking at the values shown in Fig.1. 

\noindent
A broad variation develops between approximately 10 and 80 fm/c and is associable to the pre-equilibrium $\gamma$-ray emission. In both the cases the dipolar signal stabilizes shortly thereafter with a similar recovery stage at about 120 fm/c.

\noindent
To characterize the local density evolution in the region where drift and diffusion mechanisms develop and around which the collective oscillation associated with the isovector dipolar mode takes place, we evaluated the average density as a function of time within a sphere of radius 2.5 fm centered at the center of mass (CM). The value of this radius was chosen so that in the time intervals around the instant of maximum overlap the mass of the sphere was around 12-15 units, a typical value associated with intermediate mass fragments produced at mid-rapidity.

\noindent
The local relative density, $\rho_r = \rho/\rho_0$ (with $\rho_0 = 0.16$ fm$^{-3}$), is shown in the middle panels of Fig.~2. In the interval $\Delta t \simeq 10$–$80$ fm/c, $\rho_r$ exceeds unity, reaching a maximum of approximately 1.6 and a time averaged value of about 1.3. Over the same interval, the dipolar signal performs its maximum excursion.

\noindent
The maximum of $\rho_r$ occurs around 35 fm/c, corresponding to the first inversion of the dipolar signal. The initial 10 fm/c corresponds to the approaching phase of the colliding nuclei. Beyond 80 fm/c, the relative density decreases below unity.

\noindent
To quantify the effect of the symmetry energy on the dipolar dynamics, we consider the local average force $\langle F_{\text{sym}} \rangle$ acting between the proton and neutron fluids.

\noindent
The time evolution of $\langle F_{\text{sym}} \rangle$ is shown in the lower panels. The $\gamma = 1.5$ case, exhibits large absolute values of $\langle F_{\text{sym}} \rangle$ than the softer case during the first 80 fm/c, and in agreement with the stiff character of the symmetry potential produces a smaller dipolar excursion.
Conversely the soft case instead produces a larger excursion and stronger equilibration.
It is also worth noting that $\langle F_{\text{sym}} \rangle$ takes on very small values, before changing sign, right near the first inversion point of the dipole signal.
This behavior can be more generally interpreted within the framework of the droplet model for the giant dipole resonance \cite{myers}. In this scenario, the enhanced isovector potential resulting from a stiff interaction at supra-saturation densities leads to a stiffer restoring potential expressed as a function the dipole coordinate.

\noindent
Similar behaviour is observed for other central impact parameters. 
In all cases, the asymptotic value of the dipolar signal is strongly correlated with its maximum excursion and it occurs within a time interval during which the local average density exceeds saturation.
Similar conclusions are also obtained by considering local averages defined on a sphere of radius 4.5 fm containing about 80\% of the total mass during maximum overlap, reaching a relative density of around 1.35 and an time average value in the aforementioned time interval equal to 1.2–1.25

\noindent
Therefore, these findings suggest that the sensitivity of the asymptotic dipolar signal for central collisions is predominantly linked to the symmetry energy behavior in the supra-saturation density region reached during the early compression stage of the collision.
Fig.~3  shows analogous CoMD-II results for an impact parameter $b=6.5$ fm. This value of $b$ represents approximatively the average value evaluated for the  less dissipative
events selected in this work (see next section).
In this case we observe rapid changes of the dipolar signal
between 40 and 85 fm/c corresponding to values of the local time averaged density equal to about $\rho_{0}$.  

\noindent
Finally, it is worth noting that the lower degree of isospin equilibration obtained with the stiff symmetry energy parameterization, compared with the soft one, is related to the specific dipolar  dynamics predicted by the model.

\noindent
For the mass asymmetric system under study, as discussed above, at the present incident energy, isospin equilibration proceeds through a damped pre-equilibrium dipole oscillation, which develops during the early stages of the collision around the maximum overlap of the two colliding nuclei and before the separation of the main fragments.
\noindent
For mass symmetric systems and/or in a different dynamical scenario, resulting from different model assumptions, effective interactions, or incident energies, the pre-equilibrium dipole mode could instead be over damped. In this case, the relationship between the degree of dynamical isospin equilibration and the stiffness of the symmetry energy would need to be reassessed.

\noindent
Dedicated measurements able to highlight  the pre-equilibrium dipole $\gamma$-ray spectrum could provide a direct experimental probe of the collective dynamics regime and characteristic times of the isospin-equilibration process.

\begin{figure}[h]
\centering
\includegraphics[width=10cm,clip]{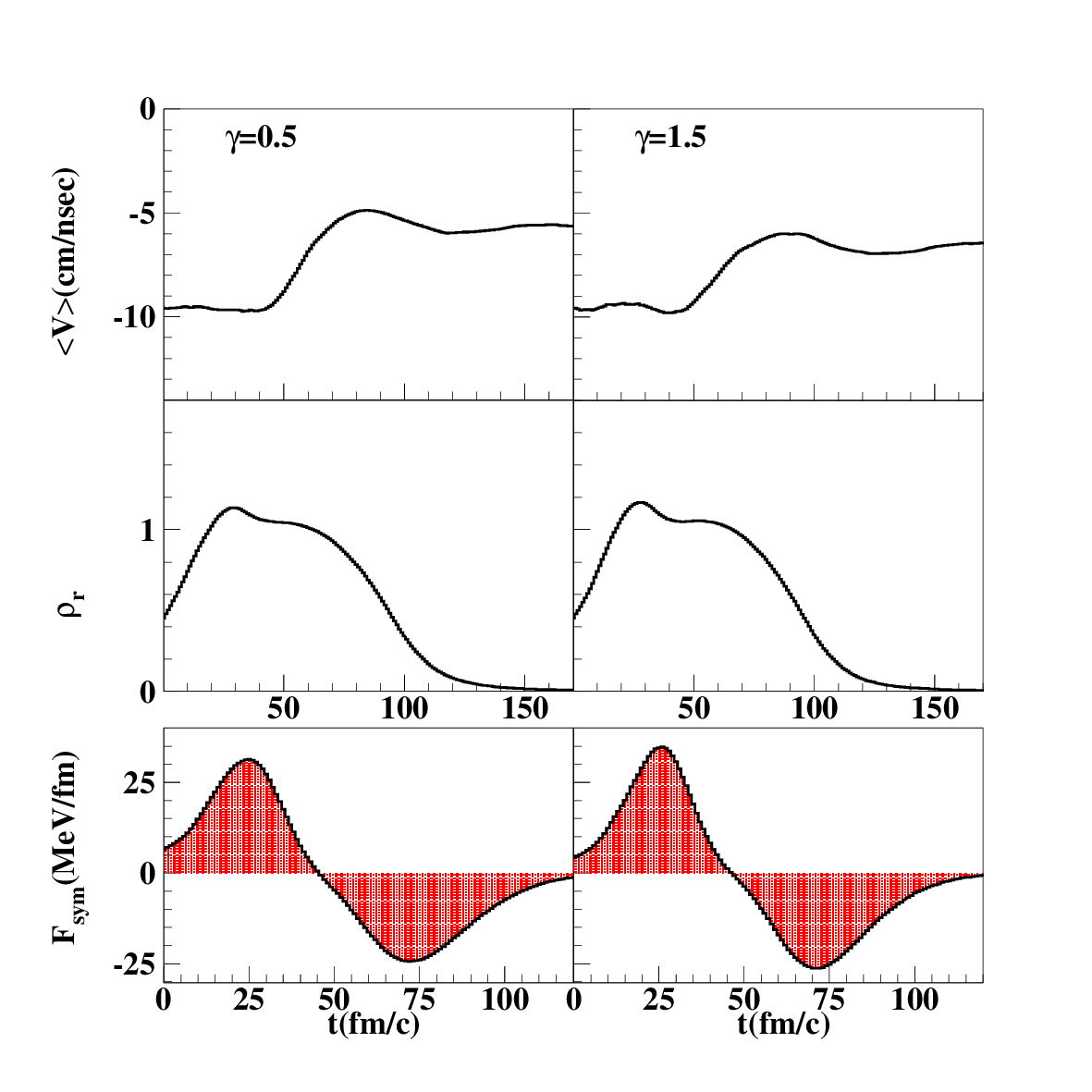}
\caption{(Colour on-line)Same as Fig.~2 but for $b=6.5$ fm.
}
\end{figure}

\section{The measurement and data analysis}

As discussed, the peculiar features of the observable $\langle \mathbf{V} \rangle$ make it suitable for providing a view of the isospin equilibration process in momentum space offering information about the simmetry energy at supra-saturation densities. For this reason, the study of this observable was extended to an experimental investigation performed with the CHIMERA multi-detector at the Laboratori Nazionali del Sud.

\noindent
The collision $^{48}$Ca+$^{27}$Al at 40 MeV/nucleon was selected. This system is rather unbalanced in charge-to-mass ratio. The use of a light target facilitates the identification of slow target-like fragments (TLF).
Fast fragments emitted in the forward direction are identified through the $\Delta E$–$E$ technique applied to the signals produced by the Si–CsI(Tl) telescopes. Time-of-flight measurements provide the fragment velocities. In particular, the timing signals allow the identification of slow fragments stopped in the silicon detectors\cite{russ}. Fast light particles are identified through the Fast–Slow correlations of the signals produced in the CsI(Tl) detectors \cite{alde}.

\noindent
 The data analysis was restricted to events for which the total charge of the identified particles is between 31 and 33 units. The reconstructed momentum is required to exceed 75\% of the nominal total value. This threshold was determined from CoMD-II plus GEMINI simulations coupled to the detector response and accounts for the maximum momentum carried away by undetected neutrons.
The stability of the results with respect to moderate variations of this threshold was verified.
Events are classified according to the degree of dissipation by evaluating the total kinetic energy in the center-of-mass frame, $TKE^{CM}$. The evaluation is performed event by event. 

\noindent
Table~1 reports the limits of the five $TKE^{CM}$ windows (IQ) considered in the present analysis.
\vskip 5pt
\begin{table}[htbp]
\centering
\caption{Limits of the $TKE^{CM}$ windows considered in the present analysis}
\begin{tabular}{|c|c|c|c|c|c|}
\hline
IQ & 1 & 2 & 3 & 4 & 5\\
\hline
$TKE^{CM}$ (MeV) & 600--450 & 450--400 & 400--325 & 325--250 & 250--100\\
\hline
\end{tabular}
\end{table}

\section{The Comparison with model calculations}
As an example, Fig.~4 shows two-dimensional plots of the fragment charge as a function of the parallel velocity $V_{z}$ along the beam axis for different $TKE^{CM}$ bins.
\begin{figure}[htbp]
\centering
\includegraphics[width=\linewidth]{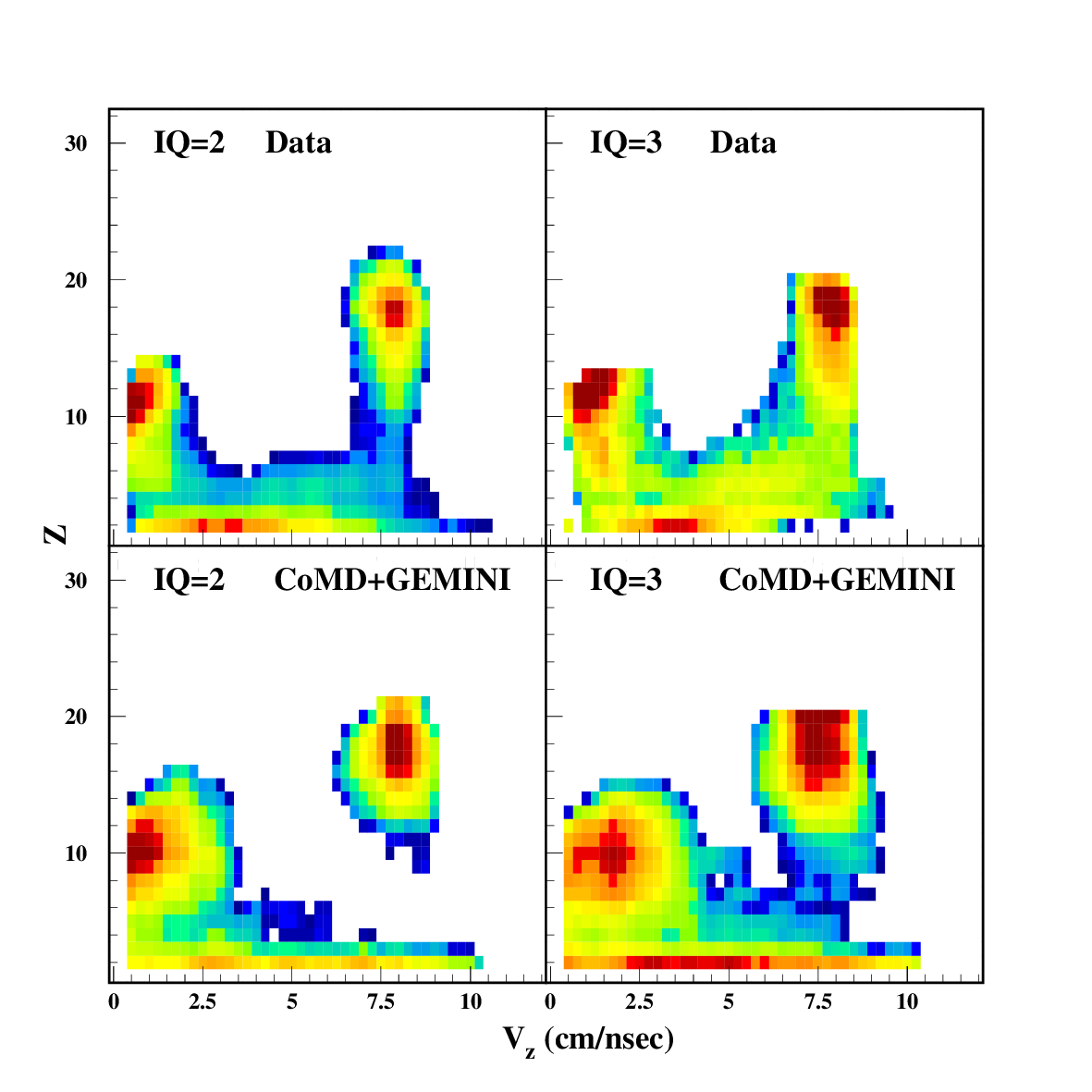}
\caption{(Colour on-line)
Fragment charge–velocity correlations plots for two dissipation windows.
Top panels: experimental data.
Bottom panels: CoMD-II + GEMINI simulations after application of the data analysis filter.
}
\end{figure}
The top panels display the experimental data, while the bottom panels show the results obtained from the CoMD-II model followed by a statistical decay stage (after a maximum time equal to 350 fm/c) obtained from the GEMINI code. The simulated events were filtered through an algorithm reproducing the same selection criteria applied to the experimental data.

\noindent
The main features of the fragment charge–velocity correlations are  reproduced by the calculations. In particular, the target-like (TLF), projectile-like (PLF), and mid-rapidity components are clearly visible.

\noindent
For each $TKE^{CM}$ window, using the identified fragments with velocity 
$V_{z}^{j}$ in well reconstructed events, the  average dipolar signal along the beam direction can be evaluated according to:
\begin{equation}
<V_{c}>=<\sum_{j=1}^{m}Z_{j}(V_{z}^{j}-V^{CM})>
\end{equation}
where the angular brackets denote the average over the selected events.
In the above expression, for each event, the center-of-mass velocity $V^{CM}$ is estimated from the velocities and masses of the identified charged fragments. 
This procedure reduces the effect on the dipolar signal of possible  uncertainties in the measured fragment velocities.

\noindent
In order to compare the measured values of $<V_{c}>$ with model predictions, CoMD-II+GEMINI calculations  were performed. The calculations were carried out for symmetry-energy values in the range 26–36 MeV, with a step of 2 MeV, and for a stiffness parameter varying between 0.5 and 1.5 with a step of 0.1 for $K_{0}$ equal to  220 MeV. Same calcutions were performed also 
for $K_{0}$ equal to 260. In this case the step for the  stiffness parameter was set to 0.2.   
The resulting events were subsequently filtered by means of an algorithm designed to simulate the multi-detector response taking also into account the same selection criteria adopted in the analysis of the experimental data.
\noindent
Examples of the comparison with model calculations are shown in Figs.~5–7. and Figs.~8–10 for $K_{0}$ equal to  220 and 260 MeV respectively.
\begin{figure}[htbp]
\centering
\includegraphics[width=\linewidth]{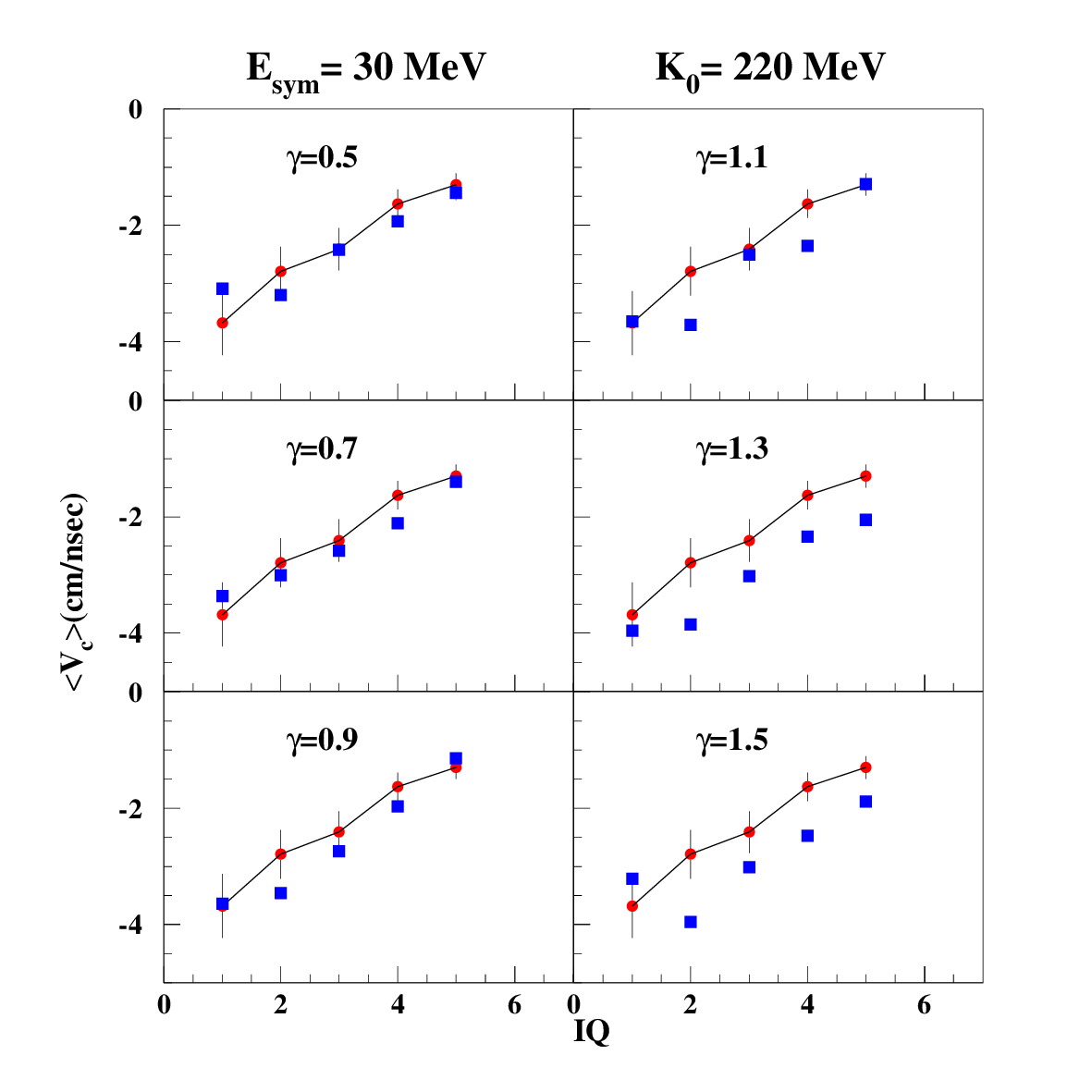}
\caption{(Colour on-line)
Comparison between experimental and calculated dipolar signals for $E_{\text{sym}}(\rho_{0})=30$ MeV and $K_{0}=220$ MeV.
The light-blue square markers represent the values of $\langle V_{c} \rangle$ obtained from CoMD-II + GEMINI calculations after application of the experimental filter for different values of the stiffness parameter $\gamma$. The associated uncertainties are smaller than the symbol size.
The experimental values are shown as red markers joined by lines; the vertical bars indicate the experimental uncertainties.}
\end{figure}

\begin{figure}[htbp]
\centering
\includegraphics[width=\linewidth]{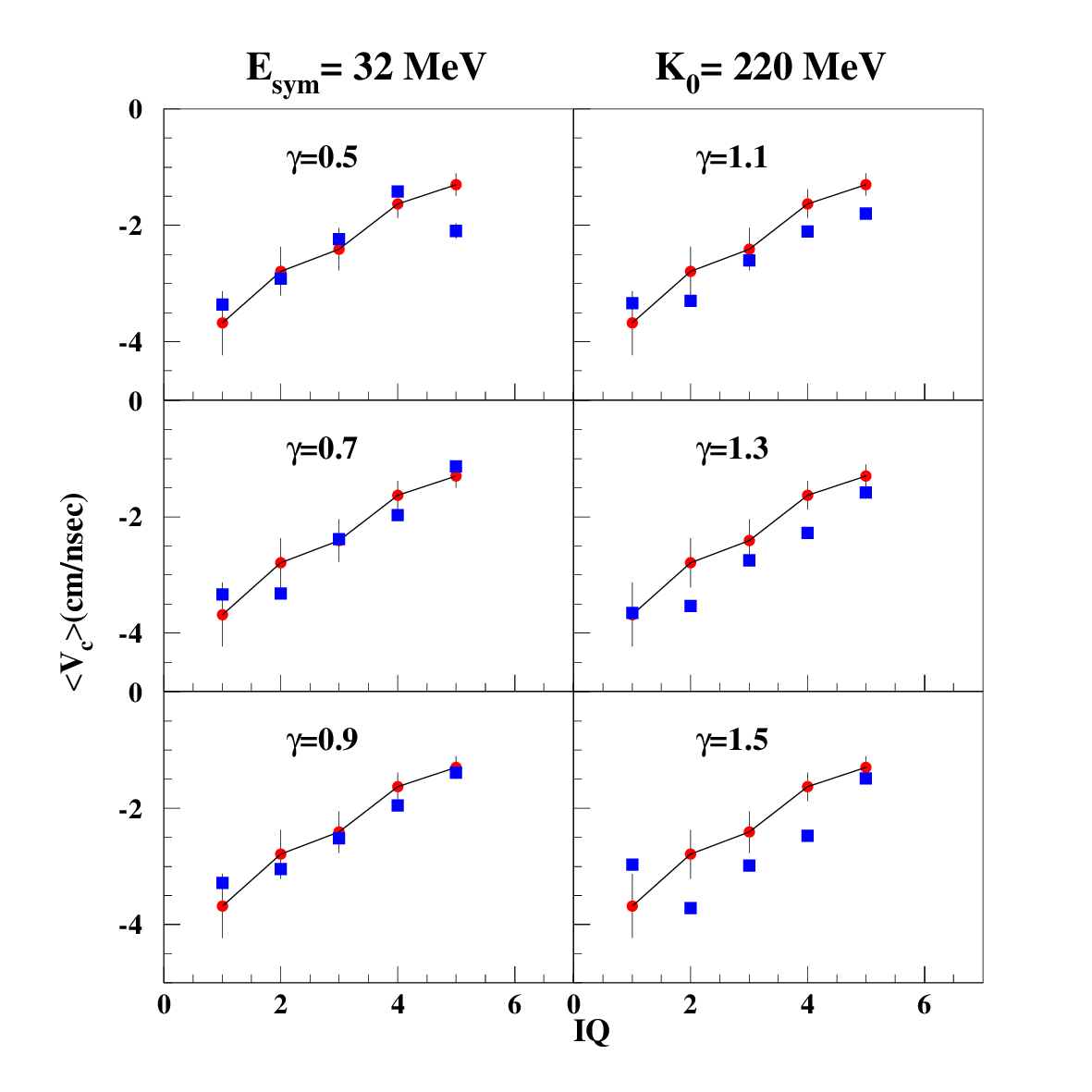}
\caption{(Colour on-line)same as Fig.~5 but for  $E_{\text{sym}}(\rho_{0})$=32 MeV.}
\end{figure}
\begin{figure}[htbp]
\centering
\includegraphics[width=\linewidth]{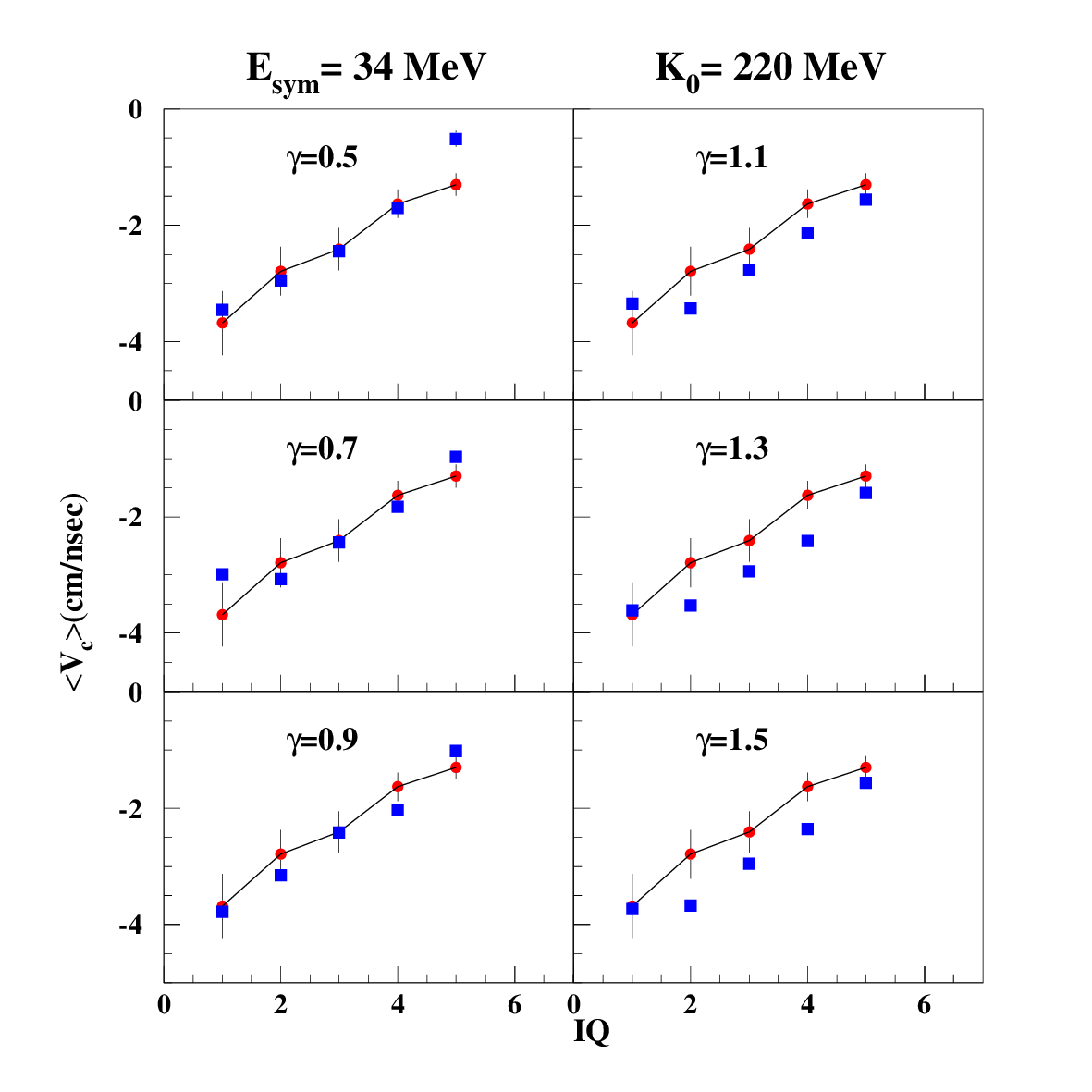}
\caption{(Colour on-line)same as Fig.~5 but for  $E_{\text{sym}}(\rho_{0})$=34 MeV.}
\end{figure}

\begin{figure}[htbp]
\centering
\includegraphics[width=\linewidth]{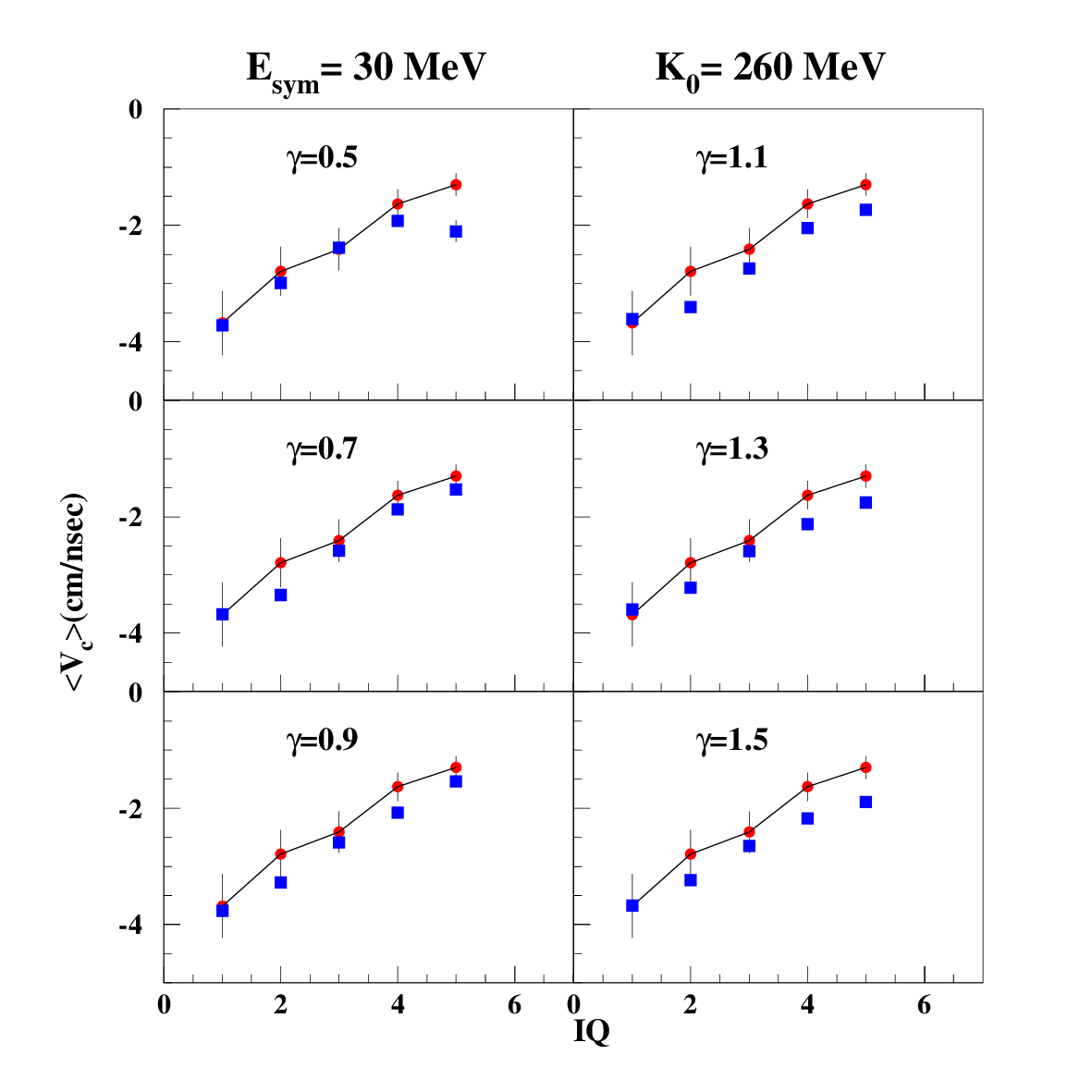}
\caption{(Colour on-line)same as Fig.~5 but for  $E_{\text{sym}}(\rho_{0})$=30 MeV and $K_{0}=260$ MeV.}
\end{figure}
\begin{figure}[htbp]
\centering
\includegraphics[width=\linewidth]{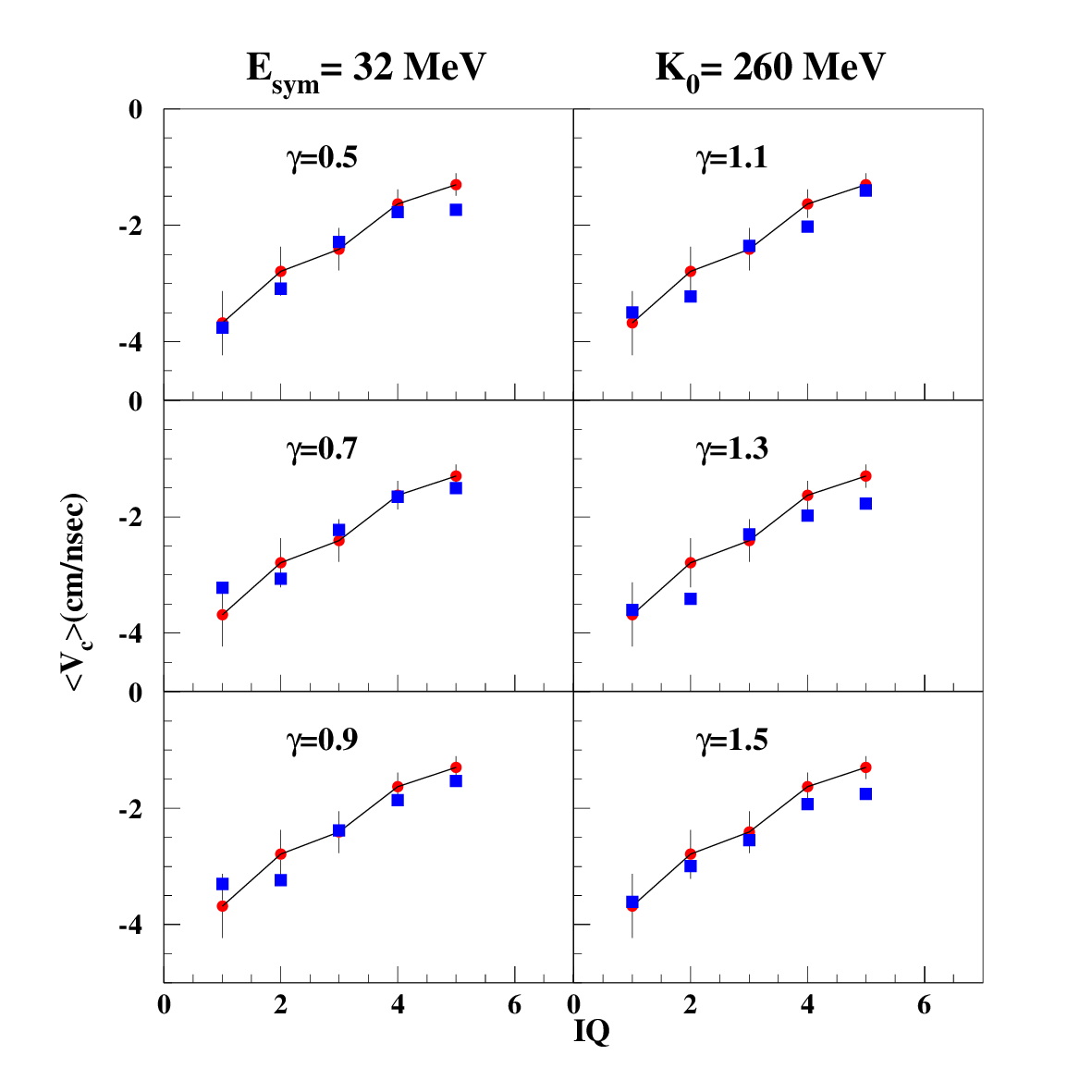}
\caption{(Colour on-line)same as Fig.~8 but for  $E_{\text{sym}}(\rho_{0})$=32 MeV.}
\end{figure}

\begin{figure}[htbp]
\centering
\includegraphics[width=\linewidth]{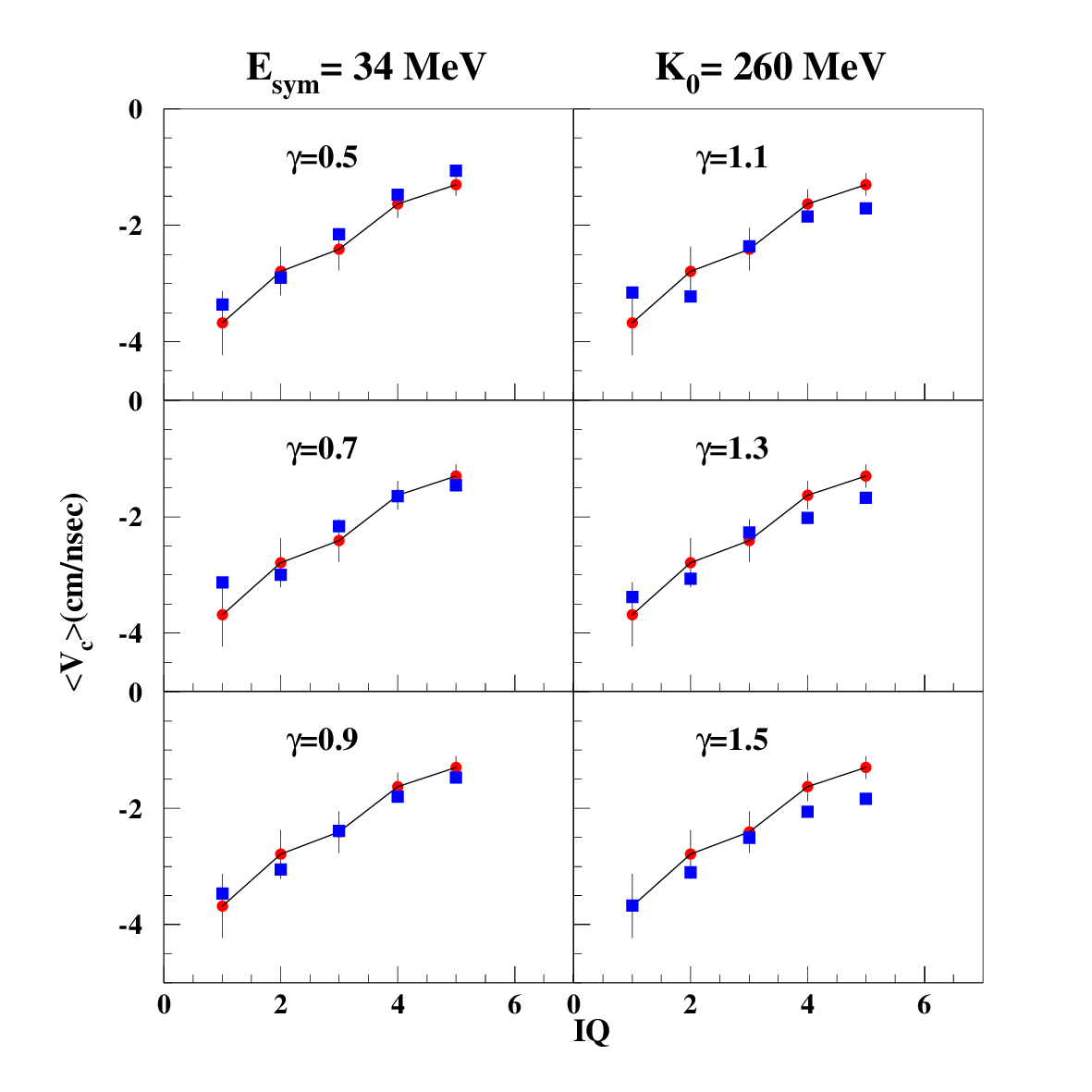}
\caption{(Colour on-line)same as Fig.~8 but for  $E_{\text{sym}}(\rho_{0})$=34 MeV.}
\end{figure}

\noindent
In these figures the experimental values of $<V_{c}>$ for the different $TKE^{CM}$ windows are reported as full red points. The error bars represent the total uncertainty on the plotted quantity, including both the statistical uncertainty and the uncertainty related to the fragment velocities. 

\noindent
The light-blue square markers correspond to the results obtained from the filtered CoMD calculations for different symmetry energies and values of the $\gamma$ parameter, as indicated. In this case the associated uncertainties are smaller than the symbol size.

\noindent
From the figures it can be seen that, with increasing dissipation (larger values of IQ), the absolute value of the measured average dipolar signal decreases. This behaviour indicates a progressive tendency toward dynamical isospin equilibration.

\noindent
The comparison with the model calculations shows that the case of the stiffest symmetry-energy displays the largest deviations from the experimental data. 
This behaviour reflects the stronger restoring force associated with a stiff symmetry energy, which tends to preserve in this case the initial isospin asymmetry of the system.

\noindent
To quantitatively analyse the comparison shown in the previous figures, for each pair of values $E_{\text{sym}}(\rho_{0}),\gamma$ we constructed a likelihood function $P_{LH}$ associated with the model results with respect to the experimental values:

\begin{equation} P_{LH}=N_{v}exp(D); \quad D=-\sum_{i=1}^{5}\dfrac{(\varepsilon^{i})^{2}}{2s_{i}^{2}}
+\alpha_{i} 
\end{equation} 
with 
\begin{equation} 
\varepsilon^{i}=\dfrac{<V^{i}_{c}>-<V^{i,exp}_{c}>}{<V^{i}_{c}>}; \quad s_{i}^{2}=\frac{ \sigma_{e,i}^{2}}{<V^{i}_{c}>^{2}}+\dfrac{<V^{i,exp}_{c}>^{2}}{<V^{i}_{c}>^{4}}\sigma_{t,i}^{2} 
\end{equation}

\noindent
In these expressions the index $i$ denotes the different IQ windows, while $\sigma_{e,i}$ and $\sigma_{t,i}$ represent the  uncertainty associated with the experimental and  theoretical  dipolar signal respectively. $N_v$ is a normalization factor.

\noindent
The second term, $\alpha_i=-\dfrac{\overline{\varepsilon}^{2}}{s_i^{2}}$ in the exponential argument $D$ is introduced to account for the effect of systematic shift   of the model calculations with respect to the experimental data.

\noindent
From the likelihood analysis on the grid of parameter $E_{\text{sym}}(\rho_{0}),\gamma$ , and $K_{0}$ using as weights the $P_{LH}$ values 
we have evaluated the average symmetry energies $\widetilde{E}_{\text{sym}}$, the average slope parameter
$\widetilde{L}$ ($L=3\rho_{0}(\dfrac{\partial E_{\text{sym}}(\rho)}{\partial \rho})_{\rho_{0}}$), the average $\widetilde{\gamma}$ 
stiffness parameter at saturation together with the symmetry energy $\widehat{E_{\text{sym}}}$ and the $\gamma$ parameter
$\widehat{\gamma}$  of the best fits with the experimental data.
The corresponding values are reported in Table~2. 

\vskip 5pt
\begin{table}[htbp]
\centering
\caption{From the likelihood analysis on the comparison between model calculations and experimental data, the average and the most probable values of the parameters characterizing the symmetry energy at the saturation density are reported.(see the text)}
\vskip 5pt

\begingroup
\renewcommand{\arraystretch}{1.2} 
\begin{tabular}{|c|c|c|c|c|c|}
\hline
$K_{0}$ (MeV) & $\widetilde{E}_{\text{sym}}$(MeV) & $\widetilde{L}$ (MeV) & $\widetilde{\gamma}$ & $\widehat{E_{\text{sym}}}$ & $\widehat{\gamma}$\\
\hline
220 & 32 $\pm$ 2 & 68 $\pm$ 10 & 0.74 $\pm$ 0.16 & 32 $\pm$ 1 & 0.9 $\pm$ 0.05\\
260 & 32 $\pm$ 2 & 73 $\pm$ 13 & 0.81 $\pm$ 0.24 & 34 $\pm$ 1 & 0.9 $\pm$ 0.1\\
\hline
\end{tabular}
\endgroup
\end{table}

\noindent
The parameters in the table, evaluated for the two values of $K_{0}$, are compatible within uncertainties.
They are also within the uncertainty in substantial agreement with the results coming by the measurements  on the neutron/proton double ratio and on the isospin diffusion transport ratios obtained in the same energy region of the present measurement \cite{tsang,lynch}.

\noindent
It is worth noting, however, that while the isospin diffusion phenomenon is usually associated by the authors with the behavior of the symmetry energy at sub-saturation densities \cite{tsang}, the behavior of the neutron elliptic flow relative to charged particles, studied in Au+Au collisions at 400 MeV/A \cite{russo1} , has been associated with a region of relative density of maximum sensitivity close to the values of 1.2–1.25 indicated in Section 2.2 of this work. In this case, the appropiate comparison with the parameters characterizing the symmetry energy, despite the widely differing incident energies, reveals a rather good agreement within the uncertainties of both experiments. 
 This especially for the $\tilde{\gamma}$ estimated in the present work for $K_{0}$ equal to 220 MeV.
 
\noindent
Finally according to the study decribed in Section 2.2 concerning the values of the time averaged density probed by the dipolar signal in semi-central collision we can reasonably estimate the following values of the symmetry a the supra-saturation densities
$\tilde{E}_{\text{sym}}(1.2\rho_{0})\simeq \tilde{E}_{\text{sym}}(\rho_{0})+\dfrac{\tilde{L}}{3}\dfrac{\Delta \rho}{\rho_{0}}=36. \pm 4$ MeV for $K_{0}=220$ MeV and $\tilde{E}_{\text{sym}}(1.2\rho_{0})\simeq 37 \pm 4$ MeV for $K_{0}=260$ MeV

\section{Summary and conclusive remarks}
A connection has been established between the excitation of the 
pre-equilibrium isovector dipole mode and the dynamics of isospin 
equilibration in the momentum space available to the heavy-ion collision, as inferred from the global 
fragment and particle production and related fluxes.
In particular, its global character makes the dipole signal particularly suitable for model-to-data tests of transport dynamics

\noindent
Trying to use the most complete information from the  CHIMERA detector, the isospin equilibration process for  
the collision $^{48}$Ca+$^{27}$Al at 

\noindent
40 MeV/nucleon has been investigated in momentum space through the analysis of the dipolar observable $\langle V_c \rangle$ evaluated in well reconstructed
events. 
The experimental results, analyzed as a function of the degree of dissipation, show a clear reduction of the dipolar signal 
with increasing dissipation, indicating a progressive approach 
toward dynamical isospin equilibration.

\noindent
A comparison with CoMD-II calculations, coupled to the statistical 
decay of the excited fragments and filtered according to the 
experimental conditions and data analysis, has been performed for different 
parameterizations of the symmetry energy. The analysis, supported 
by a likelihood-based quantitative comparison, constrains the 
symmetry energy at saturation to $E_{\text{sym}}(\rho_{0}) = 32 \pm 2$ MeV 
and yields a slope parameter $L = 68 \pm 10$ MeV
for $K_{0}=220$ MeV, similar results have been obtained for $K_{0}=260$ MeV.

\noindent
Within the CoMD-II framework, the present findings can be associated 
with the behavior of the symmetry energy at densities 
ranging from about $\rho_0$ up to $\sim 1.6\rho_0$ in the region of maximum overlap for central collision , with 
a time averaged value of about $\rho\sim 1.2\rho_0$ for semi-central collisions corresponding 
to $E_{\text{sym}}(1.2\rho_{0})=37 \pm 4$ MeV for 
$K_{0}=260$ MeV. 

\noindent
Finally we observe that a simultaneous measurement of the emitted $\gamma$ radiation is feasible and could provide, through the associated spectrum, direct access to the time evolution of the dipole signal before primary-fragment formation.

%
%
%





\section*{Acknowledgement}
We thank Dr. Carlo Rocca and Dr. Salvatore Monforte for their prompt assistance and support with the computing resources of the Physics Department of the University of Catania.

\end{document}